# Anatomy of point-contact Andreev reflection spectroscopy from the experimental point of view (review)

## Yu. G. Naidyuk[1], K. Gloos[2,3]

[1] B. Verkin Institute for Low Temperature Physics and Engineering, National Academy of Sciences of Ukraine, 47 Lenin Ave., 61103 Kharkiv, Ukraine
[2] Wihuri Physical Laboratory, Department of Physics and Astronomy, University of Turku, FIN-20014 Turku, Finland
[3] Turku University Centre for Materials and Surfaces (MatSurf), FIN-20014 Turku, Finland

*The review is devoted to the 80th anniversary of the birth of Igor Yanson, who discovered and developed the method of point contact spectroscopy, which is relevant to the main subject of this review.*

**Abstract**

We review applications of point-contact Andreev-reflection spectroscopy to study elemental superconductors, where theoretical conditions for the smallness of the point-contact size with respect to the characteristic lengths in the superconductor can be satisfied. We discuss existing theoretical models and identify new issues that have to be solved, especially when applying this method to investigate more complex superconductors. We will also demonstrate that some aspects of point-contact Andreev-reflection spectroscopy still need to be addressed even when investigating ordinary metals.

1. Introduction.
2. Early experiments and theory of normal-metal-superconductor contacts conductance.
3. Andreev-reflection (AR) experiments with metals with large coherence length.
    a) Temperature dependence of AR spectra.
    b) Magnetic field dependence of AR spectra.
4. AR in the diffusive regime.
5. What is the physical meaning of the Z parameter?
6. Origin of the broadened AR spectra.
7. Extension 1-D BTK model to 3-D.
8. Determination of spin polarization.
9. The Beloborod`ko-Omelyanchouk pair breaking model.
10. Multi band model.
11. Simultaneous study of AR and Yanson electron-phonon interaction spectra.
12. Excess current.
13. Non-AR features.
14. Conclusion.

# 1. Introduction.

Andreev reflection (AR), introduced by A.F. Andreev [1] to describe the thermal transport across a normal-metal-superconductor (N−S) interface, was used for the first time by Artemenko, Volkov and Zaitsev (AVZ) [2] to explain the nature of the so-called excess current of the current-voltage *I(V)* characteristics of superconducting (SC) point contacts (PCs). Moreover, AVZ have shown [2] that the differential conductance *dI/dV(V)* of a "dirty" (in other words – diffusive) PC displays a maximum at the SC gap value, as experimentally observed by their colleagues Gubankov and Margolin [3]. Additionally, they have also shown that the excess current is proportional to the energy gap in accord with the AVZ theory. Later, in 1982, Blonder, Tinkham, and Klapwijk (BTK) [4] proposed a "generalized semiconductor model, with the use of the Bogoliubov equations to treat the transmission and reflection of particles at the N−S interface", which allows to compute *I(V)* curves of N−S contacts ranging from metallic junctions to tunnel ones by including a barrier of arbitrary strength at the interface. The BTK equations and their modifications are widely used to extract the SC gap and other parameters of superconductors from the experimental *I(V)* curves of PCs and their derivatives. This kind of research developed eventually in PC Andreev reflection (PC AR) spectroscopy. The latter has become a popular tool for the study of unconventional superconductors, such as heavy fermions, high-$T_c$ superconductors including the recently discovered iron-based superconductors and other emergent materials. At the same time, the investigation of classical superconductors remained in the background, though it is still of direct interest to understand the AR phenomena in more details and scrutinize thoroughly both its theoretical and experimental aspects. In this review we will focus on the investigation of simple superconductors to shed more light on still open questions, which can help and be useful during the study of more complex materials.

## 2. Early experiments and theory of normal-metal-superconductor contacts conductance

As mentioned in the introduction, AVZ were the first to explain the excess current of the *I(V)* characteristics of N−S PCs utilizing AR. They found that the excess current $I_{exc}$ in N−S PC in the diffusive regime is proportional to the SC gap value: $I_{exc} = (\pi^2/4 - 1)\Delta/eR_N$, where $\Delta$ is half of the full SC gap, $R_N$ is the PC resistance in the normal state, e is the electron charge. They have also calculated *dI/dV,* which shows a maximum at $V = \Delta/e$ (or a minimum in *dV/dI,* see Fig.1). The latter was experimentally confirmed by Gubankov and Margolin [3] by investigating PCs between a pointed tantalum wire and a flat surface of copper (see Fig. 2). They have also shown that the excess current in this case is proportional to $\Delta(T)$ in agreement with the AVZ theory. Thus, such experiments with N−S PCs provide direct information of the SC gap.

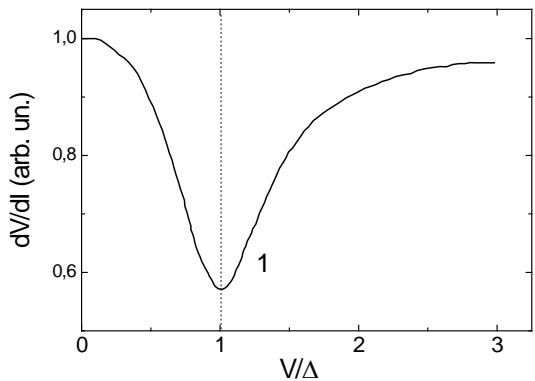

Fig.1.
Calculated differential resistance for the N−S bridge at $k_BT/\Delta = 0.1$. Adapted from [2].

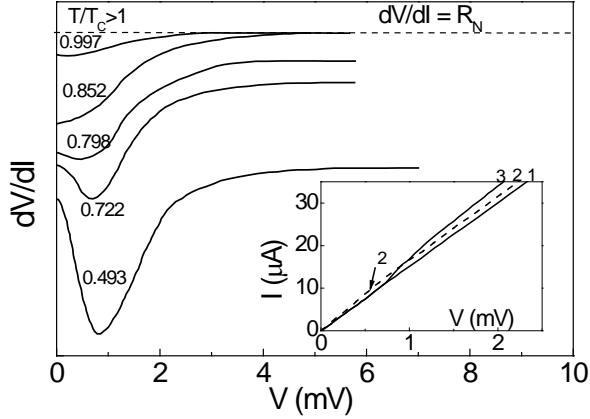

Fig. 2.
*dV/dI* for a Ta−Cu PC with $R_N$=68.5Ω at different temperatures. The curves are shifted along the resistance axis. Adapted from [3]. Inset shows *I(V)* curves at T/$T_c$>1 (1), T/$T_c$<1 (2) and T/$T_c$<<1 (3).

A few years later, BTK [4] presented their own theory for describing the *I(V)* characteristics of a clean (that is ballistic) N−S PC based on the mechanism of AR. They added a barrier of arbitrary strength, denoted as *Z*, at the N−S interface. This allowed to compute a family of *I(V)* curves ranging from the metallic PC to the tunnel junction. In the follow-up paper [5] they have applied their theory to describe *I(V)* curves of Cu−Nb PCs and found good quantitative agreement between theory and experiment. BTK mentioned that although the parameter Z plays a fundamental role in their theory, it cannot be independently determined, and must be inferred from the *I(V)* curve. They proposed [5] that Z should be thought of as a phenomenological parameter to measure the elastic scattering in the PC, whether it originates from dislocations, tunneling oxide barrier, or surface irregularities. Scattering or normal reflection can also be due to the mismatch of Fermi velocities of the contact electrodes, so that

$$Z_{eff}^2 = Z_0^{\ 2} + (1 − r)^2/(4r) \qquad (1)$$

where the ratio of the two Fermi velocities $v_{F1}$ and $v_{F2}$ of the contact metals is $r = v_{F1}/v_{F2}$ and $Z_0$ is a phenomenological parameter that contains all the other reflection mechanisms. Figure 3 shows comparison of theoretical and experimental *dV/dI* and *I(V)* curves for Nb−Cu PC for the case of barrier strength $Z = 0.65$.

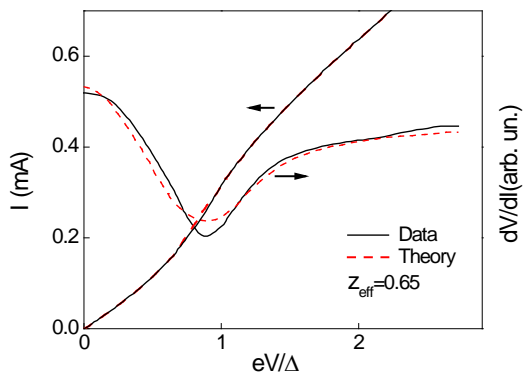

Fig. 3.
Comparison of theoretical and experimental *dV/dI* and *I(V)* curves for Nb−Cu PC for the case of barrier strength Z=0.65. Adapted from [5].

The BTK model uses a one-dimensional geometry in which, in particular, the dependence of the excitation distribution function and the reflection coefficient on the direction of the momentum is neglected. A.V. Zaitsev [6] built a more sophisticated theory, which is not based on any assumptions about the shape and the transparency of the potential barrier. Because the BTK equations for *I(V)* curves of N−S PC are reproduced in a variety of papers, we present below Zaitsev's formulae:

$$I = \frac{1}{R_N} \int_{-\infty}^{\infty} d\varepsilon B(\varepsilon) \left( th\frac{\varepsilon + eV}{2kT} - th\frac{\varepsilon - eV}{2kT} \right), \text{ where}$$

$$B(\varepsilon) = \left\langle \frac{\alpha}{(\varepsilon/\Delta)^2 + \left[1 - (\varepsilon/\Delta)^2\right]\left(2D^{-1} - 1\right)^2} \right\rangle, |\varepsilon| < \Delta \qquad (2)$$

$$B(\varepsilon) = \left\langle \frac{|\varepsilon|\alpha}{|\varepsilon| + (\varepsilon^2 - \Delta^2)^{1/2}\left(2D^{-1} - 1\right)} \right\rangle, |\varepsilon| > \Delta$$

and $\varepsilon$ is the energy with respect to the Fermi level, $D$ is the transmission coefficient, $\alpha = p_z/p_F$ with $p_z$ the momentum component perpendicular to the contact plane. These equations can be reduced to the BTK formulae [4] by taking $D^{-1} = 1 + Z^2$ and $\alpha = 1$

## 3. AR experiments with simple metals with large coherence length.

The BTK theory [4] assumes that both the energy gap and the electric potential rise to their full asymptotic values on a scale shorter than the SC coherence length $\xi$. This requires that the diameter d of the PC must be small enough $d \ll \xi$. PC AR experiments with conventional SCs that have a low critical temperature but large coherence length, like Zn ($\xi \approx 2000$ nm) and Al ($\xi \approx 1500$ nm), fulfill this condition easily.

### a) Temperature dependence of AR spectra.

Fig.4 shows the first PC AR measurements on Zn [7]. The experimental *dV/dI* curves can be perfectly described (that is fitted) by the BTK theory (see Fig.4, left inset). Both the SC gap value $\Delta$ and its temperature dependence $\Delta(T)$ (see Fig.4, right inset) agree well with the expected BCS theory of phonon mediated superconductivity. Here, for the analysis of the measured *dV/dI* curves and their temperature dependence, the modified BTK model was used, which includes the so-called lifetime or broadening parameter $\Gamma$ (for details see section 6).

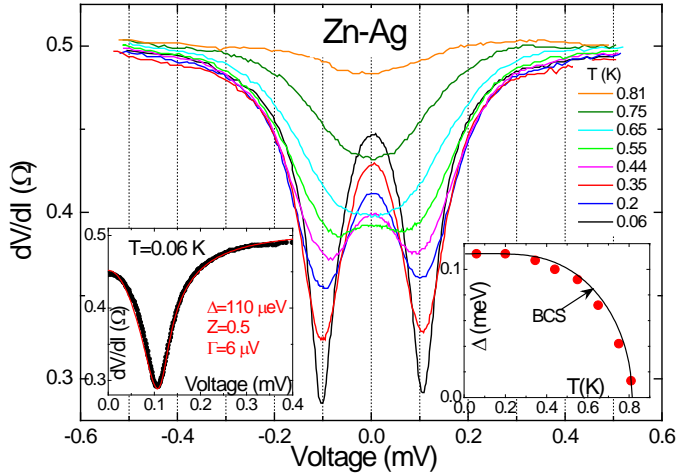

Fig.4.
Main panel: *dV/dI(V)* of a Zn−Ag PC with $R_N$=0.5 Ω at different temperatures. Left inset: Measured *dV/dI* (solid curve) at T=0.06 K and BTK fit (dashed red curve) with parameters: $\Delta$=110µeV, $\Gamma$=6µeV, and Z=0.5. Right insert: $\Delta$(T) extracted from the fit procedure. Here 2$\Delta$(0)/k$T_c$=3.1+/-0.1. Adapted from [7].

### b) Magnetic field dependence of AR spectra.

More intriguing were PC AR measurements in a magnetic field *H*. With increasing *H* either the *dV/dI* curves of Zn−Ag PC evolve smoothly towards the normal state (Fig. 5.), or the double-minimum structure abruptly disappears slightly below the bulk critical field $H_c$ (Fig. 6.).

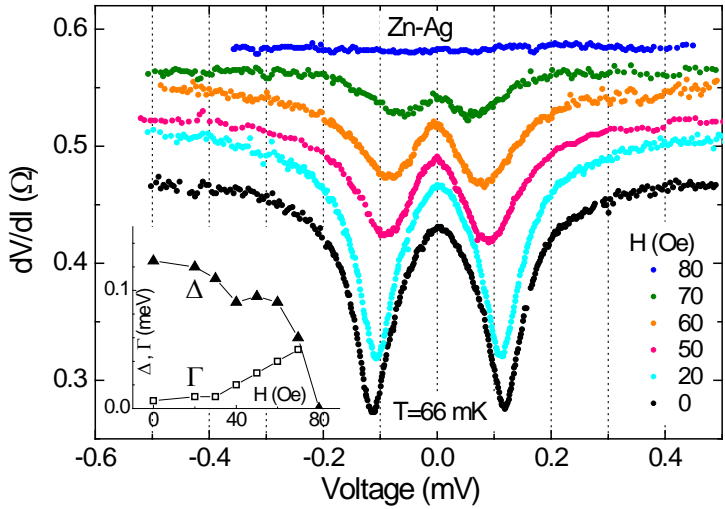

Fig.5.
*dV/dI(V)* curves of a Zn−Ag PC with $R_N$=0.47 Ω in a magnetic field at T=0.066 K. Inset shows Δ(H) and Γ(H) extracted from *dV/dI(V)* using a BTK fit. A smooth second order transition to the normal state is seen, as expected for type-II superconductors. Adapted from [7].

The original BTK theory fails to describe adequately how the *dV/dI* characteristic depends on the magnetic field. The main difference with respect to the theoretical calculations is the broadening of the experimental curves and the smaller amplitude of the AR feature. In this case the modified BTK theory, which includes the finite quasi-particle lifetime or broadening parameter $\Gamma$, describes better the magnetic field dependence of *dV/dI* curves. While fitting the zero-field *dV/dI* of the Zn contacts needs not more than a small $\Gamma<<\Delta$ [7], to fit reasonably *dV/dI* in a magnetic field often requires a significant increase of $\Gamma$ [8,9]. Additionally, AR structures often disappear at magnetic fields much higher than the bulk critical field, so that a type-I superconductor like Zn can behave at a PC like a type-II superconductor with smoothly decreasing $\Delta(H)$ (see Fig. 5) instead of a sharp transition as in Fig. 6.

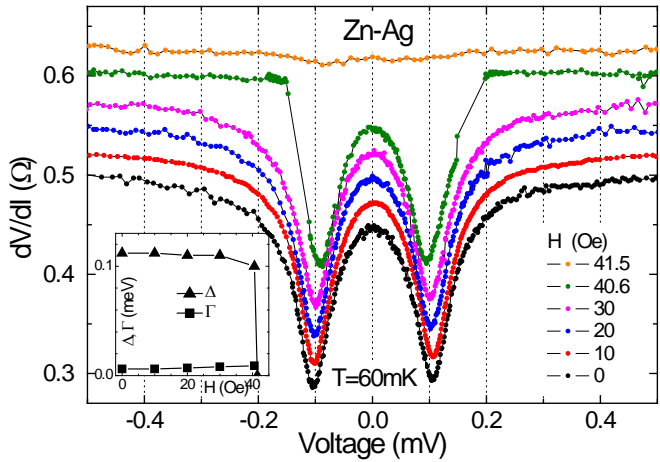

Fig.6
*dV/dI(V)* of a Zn−Ag PC with $R_N$=0.5 Ω in a magnetic field at T=0.06 K. Inset shows Δ(H) and Γ(H) extracted from the *dV/dI(V)* curves using BTK fits. The sharp first order transition to the normal state at around 40Oe is clearly seen, as expected for type-I super-conductors. Adapted from [7].

Miyoshi *et al.* [9] studied Nb−Cu PCs in a magnetic field. They suggested that the key effect in the variation of the spectra with magnetic field is due to the normal conduction channel created by the cores of Abrikosov vortices and which does not contribute to AR. Therefore the total conductance *G(V)* is the sum of the weighted normal $G_N$ and the superconducting $G_S$ channels, where $G_N$ represents the normal-state junction that does not depend on voltage while $G_S$ is described by the BTK model

$$G(V) = hG_N(V) + (1-h)G_S(V), \quad (3)$$

Both contributions are weighed depending on the fraction of the normal interface $h=H/H_{c2}$. Fig. 7 shows that by using this two-channel model the Z parameter does no longer depend on the magnetic field and the SC order parameter $\Delta$ approximates the expected parabolic dependence on $H$, supporting the suggested approach.

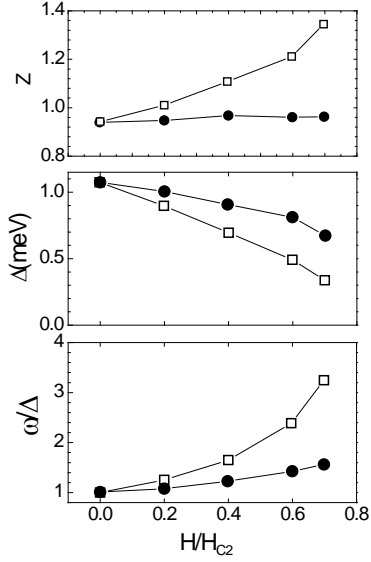

Fig.7.
Fitting parameters: SC gap $\Delta$, barrier strength $Z$ and relative spectral broadening $\omega/\Delta$[1] obtained from the experimental $dV/dI$ of Nb−Cu PC using the BTK model and neglecting the effect of the magnetic field (open symbols) and the same within the two-channel (Eq. 3) model (solid symbols). Adapted from [9].

However, the determination of the SC gap and other parameters from the $dV/dI$ spectra in a magnetic field using the standard BTK theory with BCS density of states that is broadened due to lifetime and other effects is, in principle, only a simplified approach. As discussed by Golubov and Kupriyanov [10], the SC density of states $N(\varepsilon, H)$ in the mixed state varies in space. Fig. 8 shows the density of states averaged over an elementary unit cell for several magnetic fields. In general, $N(\varepsilon, H)$ cannot be described by $N(\varepsilon, \Gamma)$ with a single lifetime parameter $\Gamma$. Therefore BTK fitting of an AR spectrum taken in a magnetic field using lifetime broadening with a single $\Gamma$ parameter is probably not a proper procedure, as discussed in [11], but requires an appropriate theoretical description which does not yet exist. Moreover, the local SC density of states in a PC will depend on the position of the pinned vortex with respect to the PC area. Therefore the obtained effective parameters of the homogeneous model and their magnetic field dependence extracted from AR spectra measured in an external magnetic field should be interpreted cautiously.

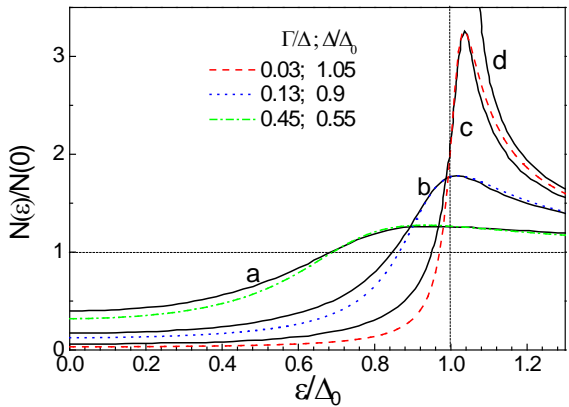

Fig. 8.
Density of states averaged over an elementary unit cell as a function of energy at $H/H_{c2}=$ (a) 0.5; (b) 0.2; (c) 0.05; (d) 0 (solid curves, numerical solution according to [10]). For comparison, $N(\varepsilon, \Gamma)$ dependencies calculated according to Eq. (4) are shown (dashed lines), which are the most similar in shape to the corresponding a, b and c curves. The parameters for the calculated curves are shown in the legend.

---

[1] Miyoshi *et al.* [9] treat the spectral broadening by calculating a convolution between between the BTK transmission coefficient based on the unaltered SC density of states and a Gaussianand a Gaussian function of width $\omega$ supposing that this generic method accounts for all sources of broadening. Since the Gaussian approximates the derivative of the Fermi-Dirac distribution, this method describes thermal broadening, leaving the density of states intact. Therefore it differs from lifetime broadening.

## 4. AR in the diffusive regime.

BTK [4] mention that the result of AVZ [2] for a micro-constriction in the dirty (that is diffusive) limit agrees with their own calculations of a ballistic junction with δ-function barrier of strength $Z \approx 0.55$. Mazin *et al.* [12] confirmed that a contact between a normal and a superconducting lead separated by a diffusive region larger than the electronic mean free path has nearly the same zero-bias resistance in the normal and in the superconducting state. They also showed that *dI/dV* spectra in the diffusive limit with $Z = 0$ and in the ballistic limit with $Z=0.55$ look similarly (see Fig. 9).

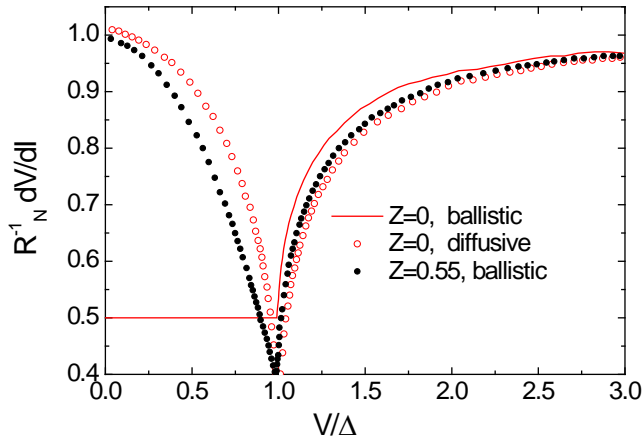

Fig. 9.
Calculated AR *dV/dI* spectra in the diffusive and the ballistic limit at T=0. Adapted from [12].

## 5. What is the physical meaning of the Z parameter?

The BTK theory describes an N−S interface with a δ-function barrier characterized by the Z parameter, see Eqs. (1). Several mechanisms can contribute to this parameter like a tunneling barrier (for example one caused by an oxide layer at the interface) or Fermi velocity mismatch.

Gloos and Tuuli [13, 14] devoted their study of N−S contacts with different normal metals (Cu, Ag, Au, Pd and Pt) and superconductors (Al, Cd, Zn, In, Sn, Ta, Nb) to elucidate if Fermi velocity mismatch influences the *Z* parameter extracted from the *dV/dI* AR spectra. Their conclusion was that the Fermi velocity mismatch does not account for the observed *Z* coefficient. Moreover, Gloos and Tuuli reported only a small variation of *Z* around 0.5 independent of the normal metal or the superconductor for PCs with different resistance (see Fig.10). Such a consistent behavior of *Z* also indicates that the interfaces usually have a negligible dielectric tunneling barrier. This excludes two of the possible normal reflection mechanisms (velocity mismatch and tunneling barrier) at least for the ordinary metals studied by Gloos and Tuuli [13, 14].

Detailed information about normal reflection can be obtained using multiple Andreev reflection at mechanically controllable break junctions. This allows to determine the transmission coefficient *T* of individual conductance channels when the PCs are very small [15].

Riquelme *et al.* [16] measured Pb−Pb junctions and found that the distribution of *T* agrees well with that of a diffusive contact. This is weird because in those contacts there is only one place where electrons can scatter, namely at the contact interface itself. This differs completely from what we usually understand by a 'diffusive contact' or diffusive transport studied by [17], which deals with 'long' channels. Riquelme *et al.*'s contacts should have been in the quantum regime and certainly not been long. Riquelme *et al.* also calculated the distribution of transmission probabilities for Pb−Pb and for Au−Au contacts for various atomic configurations at the interface. For lead contacts they found good agreement with their experiment while the gold contacts deviated a lot, tending more to the 'true' ballistic behavior. However, even with those distributions one would obtain Z values not much smaller than around 0.3.

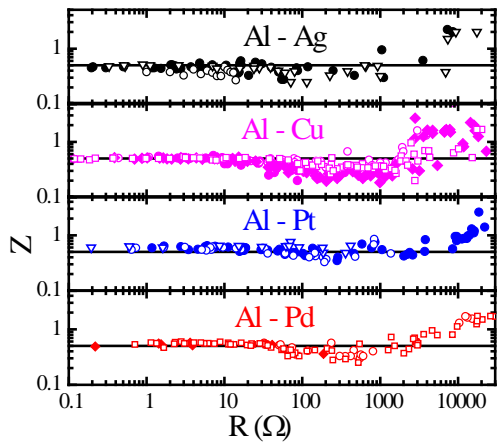

Fig.10.
$Z$ parameter versus normal PC resistance of Al in contact with the indicated metals at $T = 0.1$K. Solid lines are $Z = 0.5$ as guide to the eye. Different symbols mark different measurement series. Adapted from [13].

Makk *et al.* [18] measured (and also calculated) In−In break junctions. Their main result is how the transmission probability of the different channels (they show results for the first four channels) develops when the contact is formed. Initially when the contact is open all $T$ are zero. When the two electrodes are brought together continuously, the first channel's $T$ starts to grow, then the second one sets in and starts to grow, and so on until they saturate. If one would generalize those results then any large contact would have a certain number of channels with small $T$ which have just been formed, channels which have $T$ near 1, and those which have intermediate values. Thus they approach the $T$ distribution of a diffusive junction.

This means that a part of Z is an intrinsic property of the atomic configuration of the contacts which causes electron scattering when the lattice is disturbed. That part can reach up to around $Z = 0.55$ (it is not clear whether there this really is an upper limit). Therefore the agreement with the result for a long diffusive channel might be accidental. The theoretical results of Riquelme *et al.* [16] for Au−Au contacts show that Z can be smaller than 0.55. Whether Z can vanish completely is unclear based on those break-junction experiments and theoretical models.

Recently Gallagher *et al.* [19] have investigated AR at a structure which "consists of two superconducting strontium titanate banks flanking a nano-scale strontium titanate weak link, which is tunable at low temperatures from insulating to superconducting behaviour by a local metallic gate". In this setup the crystal lattice structure at the interface remains undisturbed and there is no dielectric barrier between the two electrodes (banks). As a result Z can be tuned by $V_G$ from zero to around 0.5.

To summarize, the main contribution of normal reflection at PCs as revealed by AR seems to origin from electron scattering at the disordered crystal lattice at the contact interface [2]. The disordered region does not need to be large since a few atomic layers are enough to produce the typical $Z = 0.5$.

## 6. Origin of the broadened Andreev reflection spectra.

Several mechanisms can broaden the AR spectra, and we can classify them according to which variables of the original BTK formalism (energy, temperature, SC gap) they affect.

---

[2] The diffusive Z = 0.55 is derived for a long channel (length >> width), this situation is unlikely for ordinary PCs. Normal reflection is naturally obtained from the disturbed crystal lattice at the contact, because the Bloch waves of the bulk electrodes are scattered there. This causes the seemingly surprising results of Riquelme *et al.* [16] and Makk *et al.* [18], that even perfect contacts have a finite normal reflection. They have shown that what looks like being diffusive comes from few atomic layers at the contact interface.

Inelastic processes shorten the Cooper pair or quasi-particle lifetime $\tau$ which in turn broadens the SC density of states $N(\varepsilon)$. Lifetime broadening and the lifetime parameter $\Gamma=\hbar/\tau$ were introduced by Dynes *et al.* [20] to describe the smeared $I(V)$ curves of tunnel junctions between strong-coupling superconductors. Plecenik *et al.* [21] incorporated the SC quasi-particle density of states

$$N(\varepsilon,\Gamma) = \operatorname{Re}\left\{ \frac{\varepsilon + i\Gamma}{\left[\left(\varepsilon + i\Gamma\right)^2 - \Delta^2\right]^{\frac{1}{2}}} \right\} \qquad (4)$$

into the BTK formalism to fit the spectra of PCs with high-$T_c$ superconductors. Wei *et al.* [22] measured AR and voltage noise of SC PCs and argued that lifetime broadening could be caused by excessive noise of the contacts. As noise source they identified two-level fluctuators in the contact region. Lifetime broadening not only broadens the spectra but strongly reduces the magnitude of the SC anomaly as well as the AR excess current.

Local heating enhances the effective contact temperature. In the BTK formalism one simply has to replace the temperature by the actual enhanced one both in the Fermi-Dirac distribution and in the SC energy gap. Local heating reduces the AR excess current, but the broadening will also depend on the applied bias voltage.

A third broadening mechanism is electrical noise. In case of white noise the effective temperature is enhanced. In the BTK formalism this leads to a broadened Fermi-Dirac distribution function without changing the SC gap. Therefore the AR excess current is not changed in contrast to lifetime broadening or local heating.

The spatial distribution of SC gap values in the contact region can be considered as another broadening mechanism (Raychaudhuri *et al.* [23], Bobrov *et al.* [24]). The conductance of such a junction is calculated by splitting it into parallel parts (analog to Myoschi's model [9] Eq. 3), each with its own gap and weight according to the SC gap distribution.

One could also consider a combination of the above mechanisms. Lifetime or spectral broadening can inform about intrinsic physical properties of the superconductor or the PC, however it does not seem to be easy to separate the different contributions.

## 7. Extension 1D BTK model to 3D.

The one-dimensional BTK theory [4] assumes that charge carriers hit the contact interface on a perpendicular trajectory. At a real junction charge carriers can arrive also from other directions, conserving the momentum component parallel to the interface, as described by Zaitsev`s model (see Section 2). Thus the transparency of the junction or its transmission coefficient will depend on the angle between the direction of the incident carriers and the normal to the interface. This is easy to understand in case of a real tunneling barrier which was studied by Mortensen *et al.* [25] as well as Daghero and Gonnelli [26]. The latter concluded that "the 3D normalized conductance practically coincides with the 1D one calculated for a properly enhanced Z value" (see Fig. 11). Therefore a fit of Andreev reflection spectra using the original 1D BTK model results in overestimating the $Z$ parameter with respect to the more appropriate 3D case.

Daghero *et al.* [27] extended the standard 1D BTK model to 2D and 3D which allows to study PC AR of anisotropic superconductors. The order parameter of those superconductors could have d-wave or other exotic symmetries or they could have a complex Fermi topology. This generalization makes the theoretical description of the spectra more difficult and multi-valued, but it is probably not necessary for the elemental superconductors that we discuss here.

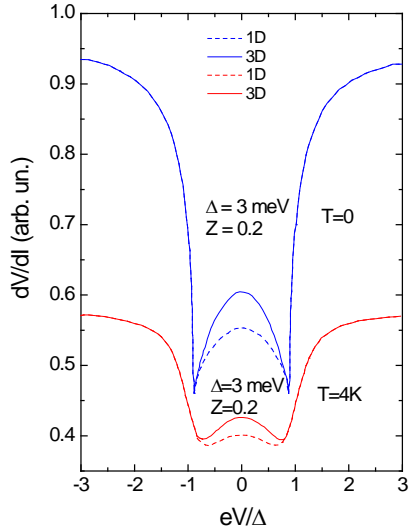

Fig.11.
Differential resistance curves *dV/dI* calculated at T = 0 and 4 K within the 1D BTK model (dashed lines) and within its 3D generalization (solid lines) using $\Delta = 3$ meV and Z = 0.2. Data are taken from [26].

## 8. Determination of spin polarization.

Initially the BTK theory [4] was mainly applied to study emergent superconductors like high-$T_c$ compounds (cuprates) [28] and heavy-fermion superconductors [29]. After a decade or so another research branch appeared, that of using AR PC spectroscopy to determine the spin polarization of ferromagnetic metals. This is possible because an electron that enters the superconductor needs another electron with opposite spin to form a Cooper pair. At a contact between a superconductor (S) and a ferromagnet (F), which has different populations of spin-up and spin-down electron bands, not all electrons find their opposite-spin counterpart. This reduces the AR probability depending on the electron spin polarization $P$. Only a fraction (1−$P$) of electrons will be Andreev reflected while the remaining fraction $P$ will be normally reflected at the F-S interface. Assuming negligible interfacial scattering (Z=0), Soulen *et al.* [30] showed that the normalized zero-bias conductance of a F−S PC is directly related to the spin-polarization $P$ via

$$R_N\,dI/dV(V{=}0){=}2(1{-}P). \qquad (5)$$

However, metal interfaces usually have Z>0, see Section 5. To take this into account, Strijkers *et al.* [31] have modified the BTK theory by splitting the conductance into two parts, one with the usual AR and weight (1-$P$) and the other one with AR probability 0 and weight $P$. Fitting the combined conductance allows to obtain the spin polarization $P$. Other authors have used slightly different versions of how to compose the two channels, for example the two-channel model of Perez-Willard *et al.* [32]. All those models have been used successfully to fit experimental AR data of a large number of PCs with ferromagnetic metals. The resulting $P$ values look reproducible and agree, in general, with expectations.

However, we note a couple of peculiarities for junctions with ferromagnetic metals (PCs with semi-metals like $CrO_2$ seem to behave differently as they show nearly full polarization [33].) First, the extracted spin polarization $P$ depends on the Z parameter of the junctions like an inverted parabola, saturating at small Z while dropping to 0 towards larger Z around 1. It is believed that the 'true' spin polarization is obtained only in the Z → 0 limit, as noted by Strijkers *et al.* [31]. Second, the typical 'true' polarization is around 0.4, more or less independently of the ferromagnet (e.g. Strijkers *et al.* [31], Naylor *et al.* [34]). Third, a number of papers report that the polarization does not change when the composition of the ferromagnetic compound or alloy is changed in order to vary its magnetism (e.g. Nakatani *et al.* [35], Naylor *et al.* [34], Osofsky *et al.* [36]). In other words there is no correlation between the extracted polarization at the point contact and the bulk magnetism. This has led to speculate that one could "fabricate an alloy with a negligible magnetization and a high transport spin polarization" (e.g. Naylor *et al.* [34]).

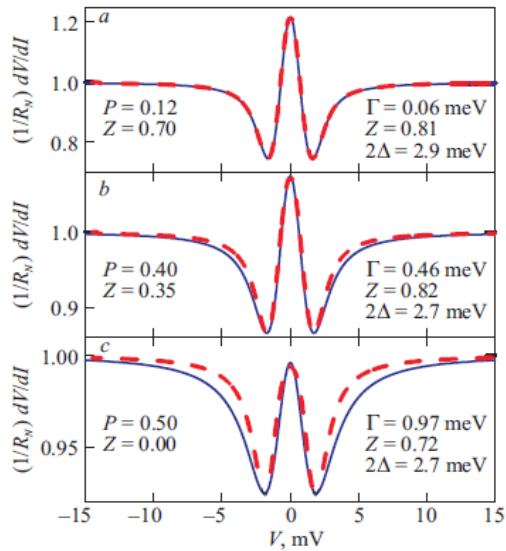

Fig.12.
Comparison between the polarization-only (red dashed lines) and lifetime-only (blue solid lines) models for contacts with (*a*) small, (*b*) medium, and (*c*) large polarization. The differential resistance *dV/dI* is normalized to the normal contact resistance $R_N$. First, the polarization-only spectra were calculated assuming the indicated *P* and *Z* at $2\Delta$=3.0 meV for niobium and *T* = 4.2K. Then the lifetime-only spectra were fitted, resulting in the indicated $\Gamma$ and *Z*. For this fitting the SC energy gap had to be slightly adjusted. After Tuuli and Gloos [40].

The main problem of using PC AR spectroscopy to measure spin polarization is to fit at least four adjustable parameters of the AR spectra: the SC gap, the normal reflection coefficient *Z*, the lifetime broadening parameter $\Gamma$ or the spectral broadening parameter $\omega$, and the polarization *P*. Different parameter combinations can fit the spectra equally well. For example, increasing *P* increases the resistance peak at zero bias and it reduces the AR excess current at large bias. A similar behavior can be achieved by increasing Z and $\Gamma$. Chalsani *et al.* [37] stated expressively that in the presence of inelastic scattering-induced pair-breaking effects "it may be impossible to distinguish between the effects of a finite spin polarization and inelastic scattering", which puts the definition of spin polarization in question. Fig. 12 compares as example theoretical AR spectra at certain spin-polarization *P* with those fitted using the modified BTK theory at *P*=0. The curves at low *P* values in the two upper panels are almost indistinguishable. As a way out of this dilemma Bugoslawsky *et al.* [38] suggested a least-mean-squares fit of the spectrum over the whole AR double-minimum structure to extract *P*. However, even this method does not really solve the problem because the differences of the calculated spectra can be quite small over a wide range of parameters (see e.g. Tuuli and Gloos [39, 40]).

In summary, PC AR spectra of S−F contacts are described well by the modified BTK theory. But it is not trivial to correctly interpret the results and to extract the genuine spin polarization.

## 9. The Beloborod'ko-Omelyanchouk pair-breaking model.

Beloborod'ko-Omelyanchouk [41] derived the electrical conductance of N–S contacts containing magnetic impurities by taking into account their pair-breaking effect. In this model spin-flip scattering of conduction electrons at the magnetic impurities with the mean free time $\tau$ has two effects on the superconductor. It changes the shape of the BCS DOS using the pair-breaking parameter $\gamma = \hbar/(\tau \Delta)$ , instead of the usual lifetime parameter $\Gamma$. And it interprets the superconducting order parameter $\Delta_{OP}$ and the energy gap $\Delta = \Delta_{OP}(1-\gamma^{2/3})^{3/2}$ as two distinct quantities. Thus, at $\gamma$=1 the superconducting order parameter is finite while the energy gap has vanished. Fig.13 shows AR spectra for different values of the $\gamma$ parameter and at a transparency *t*=0.56 corresponding to *Z*=0.89. Bobrov *et al.* [43] exploited this theory to investigate nickel-borocarbide superconductors, some of which contain magnetic rare-earth metals. The latter may behave like magnetic impurities at the disordered interface of the PC. It would be interesting to conduct similar experiments on normal metals with magnetic Kondo impurities to probe the Beloborod`ko-Omelyanchouk's theory in more detail. However, those magnetic impurities also influence the PC resistance in the normal state. For example, already 0.01% of Mn or Fe in a Cu matrix can cause a clear-cut zero-bias maximum in *dV/dI* [44]. Probably it will be very challenging to describe theoretically Andreev reflection in the presence of magnetic Kondo impurities.

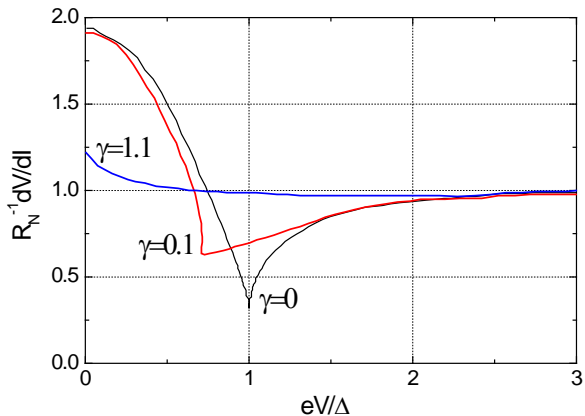

Fig.13.
Normalized differential resistance *dV/dI* for *t*=0.56 (*t* is the transparency: *t=1/(1+Z²)*) and different values of pair-breaking parameter γ. Data are taken from [42].

## 10. Multi-band model.

Multi-band PC AR spectroscopy developed rapidly after discovering superconductivity in magnesium-diboride MgB$_2$ in 2001. Different groups, see references in [26,45], measured PC AR spectra of this compound and found features of two different SC gaps. These are clearly visible in *dV/dI* as a pair of double-minimum structures because of the rather large 3:1 ratio of gap values and the weak lifetime broadening.

The fit procedure for multiband (or multigap) superconductors is simplified by supposing that each band contributes independently to the PC conductance. Thus the experimental spectra are fitted by the sum of the BTK conductance of each band with the corresponding weight factor *w*. A two-band (or two-gap) superconductor has twice the number of fit parameters plus the weight factor. With seven parameters ($\Delta_{1,2}$, $\Gamma_{1,2}$, $Z_{1,2}$, and *w*) an univocal fit is possible only if the two-gap features are visually present in *dV/dI*, like minima and/or shoulders, as is the case for MgB$_2$. One can further reduce the number of fit parameters at the cost of accuracy by assuming identical Z and/or lifetime parameters for each band.

Daghero and Gonnelli [26] reviewed thoroughly AR spectroscopy of emergent modern multiband superconductors. We searched for analogue effects in ordinary metal Zn, a most anisotropic elementary superconductor [46]. Zn has a complicated Fermi surface which includes at least three well defined sheets: the "monster", the "lens" and the "cap". In spite of the fact that the *dV/dI* AR spectra showed only a single double-minimum structure, our detailed analysis discovered "evidence for multiband superconductivity in Zn with two main gaps with the reduced gap ratio 2Δ/kT$_c$ between 3.2 and 3.7 for the small gap and between 4.2 and 5.2 for the larger one. We attribute the smaller gap to the "monster" and the larger gap to the "lens" sheet of the Fermi surface of Zn." [46]

## 11. Simultaneous study of AR and Yanson EPI spectra.

AR processes in N−S PCs are considered theoretically either in the ballistic or in the diffusive limit. Fig. 9 shows that both limits result in different shapes of *dV/dI*. On the other hand, Section 5 has revealed that contacts which should be in the ballistic limit can behave as if they are in the diffusive limit. Therefore an independent knowledge of the current regime in the PC would be useful. This could be provided by Yanson's PC spectroscopy [47]. However, among hundreds of papers exploiting PC AR spectroscopy only few have published the corresponding electron-phonon interaction (EPI) spectra which would allow to evaluate the quality of the contacts. For example, Refs [37, 48, 49] show PC EPI spectra of lead which are far from perfect, that is they have broadened phonon maxima and a large background, while the same contacts in the SC state display the clear-cut AR double-minimum structure. (The low quality of those PC EPI spectra may be related to the fact that the contacts were made in a thin-film structure.)

In our paper [46] we studied PC AR and Yanson EPI spectra on the same contacts of a superconducting Zn single crystal. Fig. 14 shows examples of AR spectra for two such PCs along with their PC EPI spectra. The first contact in Fig. 14(a) clearly shows the EPI signal while the second one in Fig.14(b) does not. Both contacts have AR spectra that are perfectly fit by the BTK theory with nearly the same Z parameter. One may conclude that (i) AR features are more robust with respect to the PC quality than the EPI spectra and (ii) for PCs close to the ballistic regime, which is confirmed by the intensity of the EPI spectra of Zn, the AR "barrier strength" Z is near the value predicted for the diffusive regime of current flow in superconducting PCs. The first (i) observation could be explained by the large coherence length of Zn compared to the size (diameter) of a PC. In this case AR reflection takes place far from the disordered interface, while information about the EPI stems from the more disturbed PC interface. As to the second (ii) observation, it looks like the characteristic length scale to describe diffusive transport differs for the two mechanisms AR and EPI scattering: the elastic electron mean free path must be compared with the PC size for the EPI features and with the coherence length in the case of AR.

In any case, Yanson PC spectroscopy has the potential to supplement AR spectroscopy to characterize the PC quality. PC spectra with EPI features proof the good contact quality.

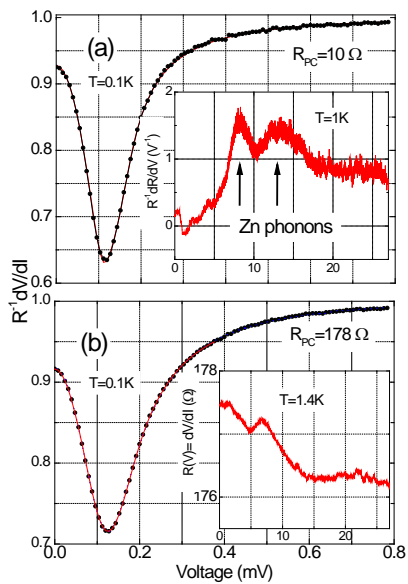

Fig.14.
Main panels: AR spectra (dots) for two PCs with different resistance along with BTK fitting (red solid lines). Upper insert shows clearly a PC EPI spectrum of Zn demonstrating a good quality of the metal in this PC, while the PC at the bottom insert shows even non metallic behavior when the differential resistance decreases with voltage bias, demonstrating a likely disordered metal state in this PC. Adapted from [46].

## 12. Excess current

Already in 1966, J. I. Pankove [50] has mentioned, that "When two oxidized superconductors are in pressure contact, a current in excess of the normal single particle tunnel current is obtained over a range of biases extending more than a decade beyond $2\Delta$". Afterwards in numerous papers dealing with the investigation of superconducting PCs the excess current $I_{exc}$ was observed. However, only in 1979 the microscopic theory by AVZ [2] explained the excess current in S−S and N−S "dirty" bridges, using the mechanism of Andreev reflection. Subsequently, the theory was extended to clean N−S junctions by A.V. Zaitsev [51]. So that

$I_{exc} = (\pi^2/4 - 1)\Delta/2eR_N$, in the dirty limit (AVZ, 1979)  (6)

and

$I_{exc} = 4\Delta/3eR_N$, in the clean limit (Zaitsev, 1980).  (7)

Khotkevich and Yanson [52] investigated the excess current of Sn−Sn contacts by characterizing the regime of current flow using Yanson's PC spectroscopy. They showed (see Fig.15) that the excess current decreases with decreasing mean free electron path as extracted from the measured intensity of PC EPI function $g_{PC}$. A similar relation between excess current and intensity of the PC EPI function was reported for N−S heterocontacts of Sn [53].

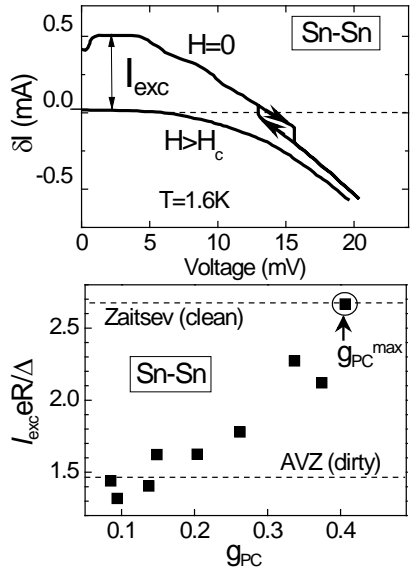

Fig. 15.
Upper panel: Deviation of the current from the ohmic behavior (dashed horizontal line) in the normal (H>H$_c$) and the superconducting (H=0) states for a Sn PC with R$_N$=2.55Ω at T=1.6K. Bottom panel: normalized excess current versus intensity of the PC EPI function $g_{PC}$ for several PCs. The horizontal dashed lines indicate the normalized excess current in the clean and the dirty limit, respectively. Data taken from [52].

Fig. 15 also shows that $I_{exc}$ decreases with increasing voltage, which can be attributed to Joule heating. In case of the hetero-contacts Sn−Cu the normal Cu electrode apparently improved the thermal coupling and reduced heating effects. As a consequence $I_{exc}$ preserved its value to higher voltages (compare Figs.15 and 16).

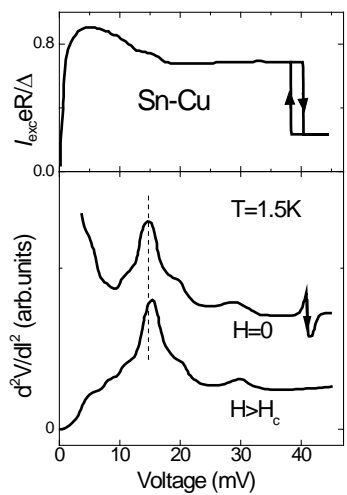

Fig.16.
Upper panel: $I_{exc}$ for PC Sn−Cu with R=8.8 Ω. Bottom panel: PC EPI spectra ($d^2V/dI^2$) of the same PC in the normal (H>H$_C$) and superconducting (H=0) state. The PC EPI spectra are similar, only a contribution from the gap structure evolves below 10 mV in the SC state. Data taken from [53].

According to the BTK theory [4], the excess current decreases with increasing scattering at the N−S interface, that is with increasing Z parameter, as shown in Fig. 17. At Z≈0.55, which characterizes the diffusive AVZ limit, $I_{exc}$ has already dropped by a factor of two. Note that I(V) curves should be measured both in the SC and normal state to extract $I_{exc}$ correctly. To reach the normal state requires applying a high magnetic field or warming the PC to above T$_c$. One can simulate the normal state $I_N(V)$ by a straight line through the origin and which runs parallel to $I(V)$ curve in SC state at high voltages as shown in Fig.18 (inset, dotted line) for a Pb–Cu PC. But suppressing the superconductivity of Pb by a magnetic field leads to a normal state $I_N(V)$ with a different slope and $I_{exc}$ that decrease continuously with a rising voltage. Note that the $dV/dI$ maximum in Fig. 18 indicates the suppression of superconductivity in the PC because $I_{exc}$ vanishes at higher voltages.

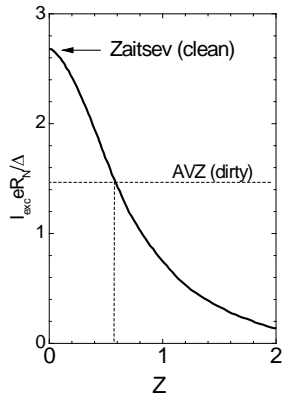

Fig.17.
Excess current as a function of barrier strength Z according data from [5].

Askerzade and Kulik [54] discussed possible non-Andreev contributions to $I_{exc}$. They assumed that a normal half-sphere develops within the superconducting half-space at the contact when the current density increases. As a result "even in the case of absence of Andreev reflection there is an excess current due to developing normal half-sphere within the SC half-space." This non-Andreev contribution depends on the ratio of electrical resistivity k=$\rho_{NM}/\rho_{SC}$ (here, $\rho_{NM}$ and $\rho_{SC}$ is the resistivity of the normal metal and the superconductor, respectively). In the case of large k (when the superconductor has a small resistivity) only the Andreev excess current remains, while in the opposite case (when the superconductor has a large resistivity) the non-Andreev excess current prevails. Thus, the destruction of the SC state in a PC by a high current density and/or heating can result in the appearance of non-Andreev contributions to the excess current. In case of a small $k$ the conductance of the PC in the SC state can exceed that in the normal state by more than a factor of two, while AR can be responsible for up to a factor of two.

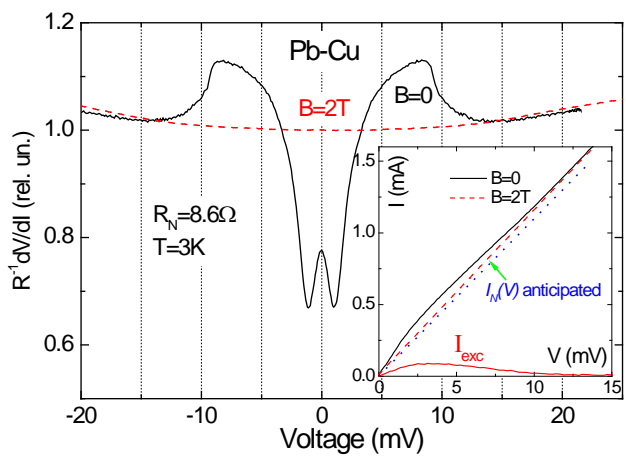

Fig.18.
$dV/dI$ spectra of a Pb–Cu PC in the superconducting (B=0, black solid curve) and the normal state (B=2T, red dashed line). Inset: $I(V)$ curves at B=0 and 2T with anticipated $I_N(V)$ (dotted blue line) for the normal state. The bottom red curve is the excess current $I_{exc}(V)= I(V, B=0)-I(V, B=2T)$.

## 13. Non-AR features.

*dV/dI* spectra have often features that are beyond the description of any known BTK model. Among those anomalies are *dV/dI* maxima (side peaks) at voltages around the SC gap like in Fig.18 and a *dV/dI* minimum at zero bias (zero-bias dip)[3].

Side peaks appear almost regularly in PCs of compounds with high normal state resistivity like high-$T_c$ [28] and heavy-fermion superconductors [55]. But they can also be observed in conventional superconductors such as Ta and Nb [56, 57]. In the latter cases the side peaks are usually connected with reaching a critical current [56] or critical (Oersted) magnetic field [57] and the subsequent transition of the PC core into the normal (resistive) state. Consequently the excess current vanishes above the side peak as shown in the inset of Fig.18.

Let us have a closer look at the side peaks (or maxima). The transition to the normal state in the PC could be a thermal effect when the inelastic electron mean free path is smaller than the PC size (diameter) and the local temperature in the PC grows with bias voltage. Even if the PC is in the ballistic or the diffusive regime at low biases, the thermal regime can develop with increasing bias [58] when the inelastic mean free path shortens sufficiently with applied voltage. This can easily take place in metals with strong electron-quasiparticle (phonon, magnon, crystal electric field etc.) interaction with a resistivity that strongly increases with temperature. In the thermal regime, when the Wiedeman-Franz law applies, the temperature in the PC and the applied voltage at $eV \gg k_BT$ are related as [58]

$$eV = 2e\sqrt{L}T = 2\pi/\sqrt{3}\ k_BT = 3.63k_BT \qquad (8)$$

where $L = \pi^2 k_B^2/3e^2$ is the Lorenz number. Thus the critical temperature Tc (and the suppression of superconductivity in the PC which manifest itself as side peak in *dV/dI*) is reached at a bias voltage of $eV_c = 3.63k_BT_c$. This almost coincides with the well-known BCS relation $2\Delta = 3.52k_BT_c$, so that $eVc \approx 2\Delta$. At a finite bath temperature $T_{bath}$ the temperature in the PC $T_{PC}$ becomes [58]

$$T^2_{PC} = T^2_{bath} + V^2/4L, \qquad (9)$$

and the bias voltage Vc required to reach Tc in the PC decreases when $T_{bath}$ increases. According to Eq.9, the resulting temperature dependence $V_c = 2L^{1/2}$ sqrt $(T^2_C - T^2_{bath})$ closely resembles the BCS behavior $\Delta(T)$. Therefore the temperature-dependent position of the side peaks could be mistakenly interpreted as a spectroscopic manifestation of the SC gap.

Westbrook and Javan [57] investigated Ta−W PCs and observed side peaks at voltages that varied linearly with the square root of the contact resistance. They interpreted the peaks as indicating the destruction of superconductivity by the self-magnetic field of the current through the contact since at the peak position the self-magnetic field reached the lower critical magnetic field $H_{c1}$ of the (type-II) superconductor

A ballistic PC with Sharvin resistance $R_S = 16\rho l/3\pi d^2$, where $\rho l = p_{F/}\ ne^2$, has a current density j which depends only on the bias voltage

$$j = V/R_sS_{PC} = V/R_s(\pi d^2/4) = 3V/4\rho l = 3en(eV)/4p_F, \ (10)$$

here $S_{PC}$ is the PC area, n is the electron density, and e is the electron charge. Large current densities of order $j \approx 10^{11}$ V [A/cm$^2$] can easily be reached for typical metals. On the other hand, the current density in a superconductor is limited by the so-called pair-breaking current density [59]

$$j \approx en_s\Delta/p_F, \qquad (11)$$

---

[3] Zero-bias minimum in *dV/dI* (maximum in *dI/dV*) characteristics for PCs of several superconductors, which is beyond BTK description, is shown in Fig.1 of Ref. [56,60]. The more probable reason for this structure is nonballistic regime with gradual suppression of superconductivity by current (and heating) with increasing of bias.

where $n_s$ is the density of SC electrons (Cooper pairs). By comparing the two expressions (10) and (11) we find that in a ballistic PC the pair-breaking current density is reached at $eV \approx \Delta$ (!). This raises the question if we can really measure AR spectra of a ballistic PC by keeping it in the SC state at $eV > \Delta$. On the other hand, in the case of a diffusive PC at $l \ll d$ (here $l$ is the elastic mean free path of electrons) the current density is lower than that of a ballistic PC by a factor $l/d$, that is $j_{diff} \approx (l/d) \, j_{ball}$. This could lead to the surprising conclusion that only in diffusive PCs one can keep the current density below its pair-breaking value at voltages comparable or even much higher than the SC gap.

The proximity effect can also cause deviations of $dV/dI$ from the BTK prediction. Strijkers *et al.* [31] considered the proximity effect with a partially reduced SC gap at the N−S interface. Their model can result in pronounced side peaks between the gap values. However, the proximity effect has two sides – a positive one by inducing SC in the normal metal and a negative one by reducing SC in the SC itself. Strijkers's model [31] describes the latter negative proximity effect. The other (positive) case leads to a S-S junction at small bias with a $dV/dI$ zero-bias dip caused by the Josephson effect and side peaks at the superconducting critical current followed by the usual AR spectrum.

## 14. Conclusion.

With the theoretical progress achieved within a few years around 1980 [2,4,6], superconducting PCs became a powerful tool to study the SC state using Andreev reflection[4]. It is nowadays known as point-contact Andreev-reflection spectroscopy. Each new class of superconductors attracts immediately this kind of investigation, and many papers are published focusing on those novel materials. We believe that a deeper understanding of the phenomena of Andreev reflection at PCs with ordinary metals reviewed in this article will be beneficial to extract useful information from and circumvent the pitfalls of the PC AR method when investigating more complex materials. With this purpose in mind we publish this review of Andreev reflection spectroscopy of conventional superconductors.

### Acknowledgments

Funding by the National Academy of Sciences of Ukraine under project Φ4-19 is gratefully acknowledged. Yu.G.N. would like to thank N.V. Gamayunova for technical help and acknowledge long term support of this activity in PC AR spectroscopy by the Alexander von Humboldt Foundation.

---

[4] Note the recent review by Klapwijk and Ryabchun [61] devoted to the 50-year-old concept of Andreev reflection and considering general aspects of Andreev reflection.